\documentclass[11pt,a4paper]{article}

\usepackage[dvips]{graphicx,graphics,color,epsfig}
\usepackage{latexsym}
\usepackage{amsmath,amsfonts,amssymb}
\usepackage{verbatim}

\begin{document}
\textwidth=150mm
\textheight=240mm

\begin{center}
\Large {\bf Fundamental physical constants: current results in search for variations and their description} 
\vskip 8mm
\large
K.A. Bronnikov, V.D. Ivashchuk, V.V. Khruschov
\vskip 5mm
\large
{{\it Russian Research Institute of Metrological Service (VNIIMS)\\ 
	 46 Ozyornaya ul., Moscow, 119361, Russia }} \\
\end{center}

\vskip 8mm

\begin{abstract}
\noindent
We consider the current results in search for and description of temporal variations of fundamental physical constants (FPCs) obtained under laboratory and astrophysical conditions. On the basis of fixed values of base constants, those FPCs have been chosen that can exhibit variations of greatest interest from the viewpoints of physics and metrology. An analysis of the current data concerning these constants is performed, and estimates of their variations on large time scales are presented. We point out the significance of studying long-term FPC variations for both practical and fundamental metrology.
\end{abstract}
\vspace*{6pt}

\noindent
Keywords: fundamental physical constant, defining constant, revised SI, variations of constants

\vskip 12mm

\section{Introduction}\label{Section_Introduction}

On May 20, 2019, the new version of the International System of Units (SI), or the reformed SI, adopted in November 2018 at the XXVI General Conference on Weights and Measures (CGPM), came into action [1]. In the reformed SI, the basic  measurement units of various quantities are redefined by fixing exact values of selected physical constants. This was done similarly to the methods of redefining the ampere in 1948, when the value
of the magnetic constant was fixed, and the meter in 1983, when the value of  the speed of light was fixed [2-4]. Instead of the term “a physical constant with fixed value”,  the 9th edition of the SI Brochure [5] introduced a new term, a defining constant. This redefinition of the base SI units will certainly contribute to improving the quality of measurements and ensuring their unity.

Taking into account the rapid development of science and modern technologies, it is necessary to continue improving the SI. One of the directions of such improvements is a search for spatial and temporal variations of the fundamental physical constants (FPCs). For the first time, the question of time dependence of the FPCs was raised by E. Milne and P. Dirac [6, 7] in 1936-1938. Dirac suggested that the ratio of the electrostatic attraction force between an electron and a proton to the gravitational attraction force between them at any given time is equal to the ratio of the age of the universe to the time during which light in vacuum travels a distance equal to the radius of an electron. This hypothesis was later extended to other FPCs, and the possible variations of FPCs began to be investigated on large spatial and temporal scales [8, 9].
This paper considers the latest results of the search for variations of the constants obtained using laboratory and astrophysical data. Some of the authors’ conclusions regarding FPC variations are also presented.

\section{Description of FFC variations based on a set of stable basic constants}\label{Section_anomal}

The XXVIth CGPM (2018) adopted a redefinition of the kilogram, ampere, kelvin and mole by fixing, without an experimental uncertainty, the values of the Planck constant $h$, the elementary charge $e$, the Boltzmann constant $k$ and the Avogadro constant $N_A$ [1]. The definition of the second has actually been preserved, but has been reformulated taking into account the exclusive role of the fixed value of the defining constant, in this case, $\Delta\nu_{Cs}$, the frequency of the hyperfine transition in the undisturbed ground state of the caesium-133 atom. The definition of the meter is based, as before, on a fixed value $c$ of the speed of light in vacuum, while the definition of the candela uses the frequency of light radiation and an energy characteristic of light, namely, the power of light radiation. The set of defining constants, making a basis for the reformed SI, must be considered, by definition, as a set of stable basic constants.

Before adoption of the new definitions of the four basic SI units at the XXVIth CGPM, numerous discussions took place regarding the advantages and disadvantages of various proposed variants of new definitions [3, 4, 10-14]. As a result, the above-mentioned set was approved. However, each of the other sets of defining constants has advantages of its own. For example, in a future redefinition of the SI unit of time, it is possible to use the set considered in [15], consisting of $h$, $c$, $k$, $N_A$, $m_e$, $\mu_0$, where $m_e$  is the electron mass, and $\mu_0$ is the magnetic constant. This set can also serve as a set of stable basic constants. 

Further on, in the description of FPC variations, we use the set $h$, $c$, $k$, $N_A$, $m_e$, $\mu_0$. In this case, in a search for variations of masses of elementary particle and atomic nuclei, we use the dimensionless quantities $\mu = m_a/m_e$, where $m_a$  is the mass of an elementary prticle or an atomic nucleus.

The coupling constants of fundamental interactions determine, to a large extent, the course of physical processes. One of the most well-known coupling constants is that of electromagnetic interactions, the fine structure constant $\alpha$ ($\alpha_e$). In the Gaussian system of units and in SI units, it is defined as
\begin{equation}
\alpha = e^2/(\hbar c);\quad
\alpha = e^2/(4\pi\epsilon_0\hbar c),
\label{eq_Umix}
\end{equation}
respectively, where $\epsilon_0$ is the electric constant, and $\hbar$  is the reduced Planck constant.

Since in the chosen system of stable basic constants $\epsilon_0=(c^2\mu_0)^{-1}$ is a stable FPC, variations of $\alpha$ are unambiguously related to variations of $e$, the elementary charge in this system of units. It should be noted that here the choice of a system of stable basic constants is connected not only with the considerations of convenience but also with the fact that the papers cited below were based on the old definitions of SI, where $\epsilon_0=\epsilon_{\rm 0\,old}$  and 
$\mu_0=\mu_{\rm 0\,old}$ are fixed, while in the reformed SI they are, figuratively speaking,
``released into free sailing'' (with the restriction  
$\epsilon_{\rm 0\,new}=(c^2\mu_{\rm 0\,new})^{-1}$) and therefore can be a source of possible variations of the fine structure constant $\alpha$. In this paper, the authors use, as the elementary charge, the old charge 
$ e \equiv e_{\rm old}$ $=e_{\rm new}(\epsilon_{\rm 0\,old}/\epsilon_{\rm 0\,new})^{1/2}$,
where the index ``new'' corresponds to the reformed SI, and the index ``old'' to the previous version of SI. 

It should be noted that the question of possible FPC variations is inextricably linked with the choice of a set of basic constants. One and the same constant may be invariable under one choice of such a set and variable under another choice. The situation with the constants $\epsilon_0$  and $\mu_0$ is an example of such a dependence. Another well-known example is that with the previously valid "astronomical" definition of a second, based on a constant period of the Earth's orbital motion, the gravitational constant $G$ was fixed, whereas with the current set of basic constants it can be variable.

To describe variations of a certain constant $\kappa$ depending on an arbitrary variable $x$, let us use the series expansion
       \begin{equation}
 \kappa(x)=\kappa(x_0)[1+(x-x_0)\lambda_0+...\;]\, ,
\label{eq_Umix1}
\end{equation}                     
where  $\lambda_0=(d \kappa(x)/d x)/\kappa(x_0)$ at $x=x_0$. In such a case,
      \begin{equation}
\Delta\kappa/\kappa \equiv (\kappa(x)-\kappa(x_0))/\kappa(x_0)=(x-x_0)\lambda_0+...
\label{eq_Umix2}
\end{equation}                    
That is, in the linear approximation, the relative variation of the constant $\kappa$ is specified by its logarithmic derivative $\lambda_0$. For example, in this approximation, the relative change in the scale factor of the Universe with time is specified by the Hubble constant
 $H=[1/a(t_0)]d a/d t(t_0)$, where $t_0$ denotes the present time.
 
 Since the FPCs are not variables in the usual meaning, it makes sense to consider long-term variations of FPCs on time scales comparable to the scale of the Universe evolution. In principle, spatial variations of the FPCs on large length scales are also acceptable. However, since we will further consider only long-term FPC variations, taking into account the general theory of relativity (GR), these variations are described with the aid of one or more scalar fields that depend on cosmological time. Such variations have been studied by many authors, for example, [6-9, 16-20].
 
 As follows from the set of stable basic constants adopted in this paper, of greatest interest are long-term variations in the mass ratios of elementary particles and atomic nuclei and the coupling constants of fundamental interactions. Long-term variation of a constant $k(t)$ will be determined by changes in the scale factor $a(t)$ of the standard Friedman-Lemaitre-Robertson-Walker metric used for the nonstationary Universe, or the cosmological redshift associated with it, $z=(1-a)/a$ if at present $a(t_0)=1$. Then,
\begin{equation}
(d k/d t)/k=-(1+z)H(z)k'/k, 
\label{eq_Umix3}
\end{equation}   
\noindent where $H(z)$ is the Hubble parameter at redshift  $z$, and  
$k'=d k(z)/d z$.  

Consider variations of the dimensionless ratio $\mu_p=m_p/m_e$,
where $m_p$  is the proton mass. 
	In the chosen system of stable basic constants, the logarithmic derivative 
	$\mu_p$ is equal to the logarithmic derivative $m_p$, but $m_p$ is mainly determined by nonperturbative effects of quantum chromodynamics (QCD) and is proportional to $\Lambda_{\rm QCD}$, where $\Lambda_{\rm QCD}$  is the scale parameter of QCD [20]. Hence, the logarithmic derivative of $m_p$ is equal to the logarithmic derivative of $\Lambda_{\rm QCD}$.
	
	Since with the selected set of stable basis constants $c$ and $\hbar$ are stable and fixed, it is always possible to pass on to a system of units in which  
	$c=\hbar=1$ . In this system, the coupling constant of strong interactions
	$\alpha_s$ in the one-loop approximation is expressed as follows:
 \begin{equation}
\alpha_s=\frac{4\pi}{(11-2n_f/3) \ln(\mu^2_R/\Lambda^2_{\rm QCD})},   \label{eq_Umix4}
\end{equation}   
\noindent                              
where $n_f$ is the number of quark flavors, and $\mu_R$ is the value of the renormalization point. 

Then the logarithmic derivative of $\Lambda_{\rm QCD}$ with respect to $z$ is proportional to the logarithmic derivative of  $\alpha_s$ with respect to $z$, with the proportionality factor 
$\ln(\mu_R/\Lambda_{\rm QCD})$  [20]. Thus we obtain a very interesting result, namely, a relationship between long-term variations of 
$\alpha_s$ and $m_p$. Moreover, since the masses of nuclei are mainly determined by the masses of nucleons, the long-term variations of the masses of nuclei are proportional to long-term variations of $\alpha_s$, which are, in turn, proportional to the long-term variations of $\Lambda_{\rm QCD}$.

For electromagnetic and weak interactions, there is no scale parameter like 
$\Lambda_{\rm QCD}$, however, there are models in which variations of the coupling constants of these interactions are much smaller than the variations of $\alpha_s$ [19, 20]. Thus, of greatest interest among the constants of the Standard Model (SM) of electromagnetic, weak and strong interactions, which is the modern quantum theory of fundamental interactions, is a search for variations of the two constants 
$\alpha_s$ and $\mu_p$.

The equations of the gravitation theory contain two FPCs, the Newtonian gravitational constant $G$ and the cosmological constant $\Lambda$: 
 \begin{equation}
R_{\mu\nu}-g_{\mu\nu}R/2=8\pi GT_{\mu\nu}/c^4-\Lambda g_{\mu\nu}, 
\label{eq_Umix5}
\end{equation}                           
\noindent where $R_{\mu\nu}$ is the Ricci tensor constructed from the metric tensor $g_{\mu\nu}$ and its derivatives; $R$ is the Ricci scalar, and $T_{\mu\nu}$ is the energy-momentum tensor of matter.

The nature of the cosmological constant remains ultimately unclear. It can be written, by analogy with the first term on the right-hand side of equations (6), as
 $ \Lambda=(8\pi/c^4) G\rho_{\rm vac} +\Lambda_0$, $\rho_{\rm vac}$ is the vacuum energy density.

The vacuum energy density is contributed by both zero modes of quantum fields and by the condensates arising at spontaneous violation of various symmetries. The values of these contributions exceed the observed vacuum energy density by many orders of magnitude. To cope with this situation, an additional term $\Lambda_0$ is arbitrarily introduced. Note that although variations of the effective cosmological constant are considered in a number of theoretical models, their estimates are largely model-dependent. At the same time, within the framework of GR, the parameter $\Lambda$ is strictly constant. Due to these circumstances, variations of $\Lambda$ will not be further discussed.

\section{Experimental constraints on relative variation rates of  $\alpha$, $\mu_p$, $G$}
\label{Section_OscillationModel}

As follows from Eq. (2), the relative variation of an FPC in the linear approximation depends on its logarithmic derivative. Let us therefore present the latest experimental constraints on the quantities $\dot{\alpha}/\alpha$, 
$\dot{\mu_p}/\mu_p$, 
$\dot{G}/G$.  The upper laboratory constraints on $\dot{\alpha}/\alpha$, obtained by comparison of optical clocks, give [21, 22]  
$|\dot{\alpha}/\alpha|\le 2\cdot 10^{-17}\,\rm yr^{-1}$, and astrophysical observations of extragalactic sources give practically the same constraint [23]. At the same time, a number of studies based on a comparative analysis of the absorption spectra of quasars located in different directions on the celestial sphere have claimed the detection of not only temporal but also spatial variations of $\alpha$ (the so-called Australian dipole [24]). A few theoretical models have also been proposed to explain such variations, for example, in the framework of a nonlinear multidimensional theory of gravity [25-27]. However, these results have not been independently confirmed by now, and the necessity of additional observations and their analysis using an improved methodology is recognized [28].

The strongest constraint on  $\dot{\mu_p}/\mu_p$ have been obtained by now by comparison of ytterbium optical clocks with caesium microwave clocks [29]: 
$\dot{\mu_p}/\mu_p \le (-5.3\pm 6.5)\cdot 10^{-17}\rm  yr^{-1}$. 

As to the best upper bounds on  $\dot{G}/G$, the experimental constraints are here somewhat less restrictive. Let us present the constraints obtained in the recent years. According to [30], by estimates obtained by tracking spacecrafts on Mars, it holds  $|\dot{G}/G|\le 10^{-13} \rm yr^{-1}$. 

In Ref. [31], based on the analysis of more than 635 000 position observations of planets and spacecrafts over a period of 1961 to 2010, the following estimate has been obtained: 
$-4.2\cdot 10^{-14} <\dot{G}/G< 7.5\cdot 10^{-14}\,\rm yr^{-1}$ at a confidence level of 95 \%. 
This estimate was recently improved by the same group [32] based on estimating the change rate of the Sun's gravitational parameter: $d(GM_{\odot})/d t$ ($M_{\odot}$ is the Sun's mass),  the numerical planetary ephemeris EPM2019 and an estimate of  $\dot{M_{\odot}}/M_{\odot}$ (matter accretion by the Sun minus radiation and solar wind losses). It was obtained that  
$-2.9\cdot 10^{-14} <\dot{G}/G< 4.6\cdot 10^{-14}\,\rm yr^{-1}$ at a confidence level of 99,7 \%. 
Ref. [33], a paper published at the end of 2020, presents the results of an analysis of long-term experiments with laser ranging of corner reflectors installed on the Moon: 
$\dot{G}/G \le (-5.0\pm 9.6)\cdot 10^{-15}\,\rm yr^{-1}$. Lastly, quite recently the following result was reported [34]: $\dot{G}/G \le (-0.110\pm 2.449)\cdot 10^{-15}\,\rm yr^{-1}$, based on astrophysical observations of the white dwarf G191-B2B, which seems to be the most stringent restriction on possible variations of the constant 
$G$ known to date. Recalling that the Hubble parameter, characterizing the expansion rate of the Universe, is estimated as  $H\approx 7.1\cdot 10^{-11}\,\rm yr^{-1}$  (note that the modern data on $H$ have a significant scatter, see, for instance, [35, 36]), it can be concluded that the gravitational constant in the modern era can change at most by four orders of magnitude slower than the expansion of the Universe.

There are also known constraints on the time variations of $G$ over long periods of the cosmological evolution. Thus, a combination of various cosmological tests leads to the conclusion [37] that at the recombination epoch, i.e., at the transition from an electromagnetic plasma to neutral matter, the value of $G$ could differ from its modern value by no more than 3 \% (at a confidence level of 68 \%).

These results impose strong restrictions on the parameters of many theories of gravity alternative to GR. For example, the well-known scalar-tensor theories have the Einstein limit as their parameter $\omega$, characterizing the scalar field coupling with curvature, tends to infinity; according to the totality of measurements in the Solar system, this parameter should currently be not less than 40,000 [38]. The recently formulated scalar-metric theory of gravity [39], which gained some success, has turned out to be incompatible with observations precisely because of a too rapid change in the effective gravitational constant predicted by this theory [40].

It should be noted that, despite quite strong constraints on its time variations, the gravitational constant $G$ remains to be one of the least accurately measured constants. The CODATA (1918) recommended value [41] is 
$G=6.67430(15)\cdot 10^{-11}\,\rm m^3 kg^{-1}s^{-2}$
  with a confidence level of 68 \% and the relative standard uncertainty 
  $2,2\cdot 10^{-5}$. Increasing the measurement accuracy of $G$ is one of the problems of modern physics and metrology.

\section{Conclusion}
\label{Section_Conclusion}

In this paper, based on the choice of a set of stable base constants, we have considered the possible long-term variations of a number of FPCs. Of greatest interest are variations of the coupling constant $\alpha$ of the electromagnetic interactions, the electron to proton mass ratio  $\mu_p=m_p/m_e$, and the gravitational constant $G$. The latest experimental limits on long-term variations of these constants are presented, from which it follows that a study of these variations is of interest for the needs of astrophysics of very distant objects as well as for the construction of a grand unification theory of fundamental interactions. As noted in the previous section, modern experimental constraints on the relative change rate of the gravitational constant allow us to exclude some alternative theories of gravity, for example, such as some generalizations of the Brans-Dicke model, a number of models of string origin [8] and the scalar-metric theory of gravity [39].
Still in the foreseeable future, the possible influence of the effects of long-term FPC variations may be ignored while solving problems of practical metrology. However, their possibility is of basic importance in fundamental metrology and should be reflected in training courses for metrologists, both for the formation of students' ideas on the basics of modern physics and because, as a result of the progress of measurement technology, these effects may begin to exert influence on the measurement results.

\end{document}